\documentclass{aastex}
\usepackage{emulateapj5}

\newcommand{\xte}{{\it RXTE}}

\newcommand{\epcs}{{\rm ergs\,cm^{-2}\,s^{-1}}}
\newcommand{\epc}{{\rm ergs\,cm^{-2}}}
\newcommand{\cts}{{\rm count\,s^{-1}}}
\newcommand{\gin}{{\it Ginga}}

\newcommand{\osoe}{{\it OSO-8}}
\newcommand{\sax}{{\it BeppoSAX}}
\newcommand{\asca}{{\it ASCA}}

\newcommand{\exo}{{\it EXOSAT}}
\newcommand{\src}{GS~1826$-$24}

\slugcomment{Accepted by \apj}

\shortauthors{Galloway et al.}
\shorttitle{X-ray bursts from \src}

\begin{document}

\title{
Periodic Thermonuclear X-ray Bursts from \src\ and
the Fuel Composition as a Function of Accretion Rate
}

\author{Duncan K. Galloway\altaffilmark{1},
   Andrew Cumming\altaffilmark{2,3},
   Erik Kuulkers\altaffilmark{4},
   Lars Bildsten\altaffilmark{5},
   Deepto Chakrabarty\altaffilmark{1,6,7}, and 
   Richard E. Rothschild\altaffilmark{8}}

\altaffiltext{1}{ Center for Space Research,
   Massachusetts Institute of Technology, Cambridge, MA 02139, email:
   duncan@space.mit.edu}
\altaffiltext{2}{Dept.~of Astronomy and Astrophysics, University of
California, Santa Cruz, CA 95064; email: cumming@ucolick.org}
\altaffiltext{3}{Hubble Fellow}
\altaffiltext{4}{ESA-ESTEC Science Operations \& Data Systems Division
SCI-SDG, Keplerlaan 1, 2201 AZ Noordwijk, Netherlands, email:
ekuulker@rssd.esa.int}
\altaffiltext{5}{Kavli Institute for Theoretical Physics and
Department of Physics, University
of California, Santa Barbara, CA 93106; email:
bildsten@kitp.ucsb.edu}
\altaffiltext{6}{also Department of Physics, Massachusetts Institute of
Technology, email: deepto@space.mit.edu}
\altaffiltext{7}{Alfred P. Sloan Research Fellow}
\altaffiltext{8}{Center for Astrophysics and Space Science, University of
California, San Diego, La Jolla, CA 92093, email:
rrothschild@ucsd.edu
\addtocounter{footnote}{-8} }

\begin{abstract}
We analyze 24 type I X-ray bursts from \src\ observed by the {\it
Rossi X-ray Timing Explorer} between 1997 November and 2002 July. The
bursts observed between 1997--98 were consistent with a stable
recurrence time of $5.74\pm 0.13$~hr.  The persistent intensity of
\src\ increased by 36\% between 1997--2000, by which time the burst
interval had decreased to $4.10\pm 0.08$~hr.  In 2002 July the
recurrence time was shorter again, at $3.56\pm0.03$~hr.  The bursts
within each epoch had remarkably identical lightcurves over the full
$\approx 150$~s burst duration; both the initial decay timescale from
the peak, and the burst fluence, increased slightly with the rise in
persistent flux.  The decrease in the burst recurrence time was
proportional to 
$\dot M^{-1.05\pm0.02}$ (assuming that $\dot M$ is linearly
proportional to the X-ray flux), so that the ratio $\alpha$ between
the integrated persistent and burst fluxes was inversely correlated
with $\dot{M}$. The average value of $\alpha$ was
$41.7\pm1.6$.
Both the
$\alpha$ value, and the long burst durations indicate that the
hydrogen is burning during the burst via the rapid-proton (rp)
process. The variation in $\alpha$ with $\dot M$ implies that hydrogen
is burning stably between bursts, requiring solar metallicity ($Z\sim
0.02$) in the accreted layer. We show that solar metallicity ignition
models naturally reproduce the observed burst energies, but do not
match the observed variations in recurrence time and burst
fluence. Low metallicity models ($Z\sim 0.001$) reproduce the observed
trends in recurrence time and fluence, but are ruled out by the
variation in $\alpha$. We discuss possible explanations, including
extra heating between bursts, or that the fraction of the neutron star
covered by the accreted fuel increases with $\dot M$.
\end{abstract}


\section{Introduction}

Studies of recurrence times and energetics of type I X-ray bursts in
low-mass X-ray binaries (LMXBs) led to one of the most compelling
arguments for their origin as unstable thermonuclear burning of
accreted hydrogen and helium \cite[e.g.][]{sb03}. Long duration
broadband X-ray observations with satellites in high orbits such as
\exo\/ allowed high-precision measurements of source fluxes and
uninterrupted trains of X-ray bursts
\cite[e.g.][]{gottwald86,langmeier87}. In these sequences of bursts,
the ratio of persistent fluence between bursts $E_{\rm p}$ to burst
fluence $E_{\rm b}$,
\begin{equation}\label{ratio}
\alpha\equiv { E_{\rm p}\over E_{\rm b} }=\frac{\int_0^{\Delta t}
      F_{\rm p}\,dt}{\int_0^{\Delta t} F_{\rm b}\,dt},
\end{equation}
where $F_{\rm p}$ is the persistent flux from accretion, $F_{\rm b}$
is the burst flux, and $\Delta t$ is the interval from the beginning
of one burst to the next, was found to be $\sim 10$--$100$. This is
consistent with the energy release in bursts coming from nuclear
burning, which gives $\approx 1.6\ {\rm MeV}$ ($5\ {\rm MeV}$) per
nucleon when burning pure helium (solar abundance material) to heavy
elements, compared to $\approx 200\ {\rm MeV}$ per nucleon from
gravitational energy release due to accretion. The general agreement
of measured $\alpha$ values with this picture was an important
confirmation of the thermonuclear burst model.

Detailed comparisons of observations and theory have had mixed
success, however, particularly attempts to reconcile the observed trends in
burst behavior with accretion rate with those expected from theory
\cite[]{fuji87,vppl88,bil00,corn03a,cumming03}. Unfortunately,
observational studies are now more difficult because most of the
active broadband X-ray observatories are in low Earth
orbits. Interruptions in the observations due to occultations and
passages through localised high-background regions occur at the
satellite orbital period, which is of the same order as the typical
recurrence time for bursts.

Despite such obstacles, \src\/
(also known as Ginga 1826$-$238;
$l=9\fdg27$, $b=-6\fdg09$) is an ideal subject for studying X-ray
bursts due to its bright, regular and frequent bursts. The source was
discovered as a new transient by \gin\/ \cite[]{tanaka89}.
Conclusive evidence of the presence of a neutron star was obtained
with the \sax\/ detection of thermonuclear bursts \cite[]{ubert97},
although this source may also have been the origin of X-ray bursts
observed much earlier by \osoe\/ \cite[]{becker76b}.
Optical photometry of the $V=19$ counterpart revealed a 2.1~hr
modulation, as well as optical bursts \cite[]{homer98}. The delay time
measured between the X-ray and optical bursts is consistent with the
binary separation for a 2.1~hr orbit \cite[see also][]{kong00}.  Based
on optical measurements, the distance, $d$, to the source is at least
4~kpc \cite[]{barret95}. An upper limit of 8~kpc has been derived from
the peak fluxes of bursts measured by \sax, \asca\/ and \xte\/
\cite[]{zand99,kong00}, which would position the source just outside
the Galactic bulge. The global accretion rate inferred from the observed flux
is then
$\dot M\approx 10^{-9} M_\odot {\rm yr^{-1}}(F_{\rm p}/3\times
10^{-9}\ \epcs)(d/6\ {\rm kpc})^2$, about a factor of ten higher than
the expected time-averaged accretion rate for a NS orbiting a main
sequence star in a 2.1~hr orbit
\cite[]{vvdh95}. Such a discrepancy is not surprising for a transient
accretor.

Extensive \sax\ Wide Field Camera (WFC) observations between 1996 August
and 1998 October revealed around 70 bursts, which recurred on an average
interval of $5.76\pm 0.62$~hr \cite[leading to the pseudonym ``the Clocked
burster'';][]{clock99}.
Analysis of the entire WFC sample of 260 bursts revealed that the
source consistently exhibits approximately periodic bursts with a
recurrence time which decreases significantly as the persistent flux
increases \cite[]{corn03a}.
Such regularity of bursts over a long time span is not observed in other
sources and implies a remarkably stable accretion rate, as well as
complete burning of the accreted fuel and a constant fuel covering
fraction. 

Based on the decay time scale of the bursts observed by \xte\/
\cite[]{kong00} and the recurrence times, \cite{bil00} suggested that
the bursts must involve mixed hydrogen/helium (H/He) burning. In this
scenario, the initial helium flash generates seed nuclei for hydrogen
burning, which gives rise to an additional energy release and also
allows an active rapid proton (rp) process of \cite{ww81}.  This
process burns hydrogen by successive proton capture reactions on seed
nuclei that subsequently $\beta$-decay. These reactions move the seed
nuclei up the proton-rich side of the periodic table (much like the
r-process which occurs by neutron captures on the neutron-rich side of
the periodic table). Theoretical work indicates that the end point of
the rp-process is set either by the complete consumption of the
available hydrogen \cite[]{hsh83,remb97,schatz98,koike99} or by
reaching the closed SnSbTe cycle found by \cite{schatz01}.
Uncertainties still remain in the actual total time to burn the
matter, since the $\beta$-decay lifetimes of many of these proton-rich
nuclei are not known experimentally. Thus, \src\ is an ideal candidate
to test, and provide input for, our current theoretical understanding
of the role of the rp-process in thermonuclear X-ray bursts.

We present new measurements of the burst recurrence time in \src\/ with
the {\it Rossi X-ray Timing Explorer} (\xte). We first compare the
properties of different bursts to highlight their unusually uniform
properties.
We examine the long-term trend of burst recurrence time as a function of
persistent flux, and test for variations in the $\alpha$-value within
different epochs.  By comparing to theoretical models of type I bursts, we
are able to constrain the metallicity of the accreting material, and the
fraction of the neutron star's surface over which accretion takes place.

\section{Observations}

We analysed observations of \src\/ by the All Sky Monitor
\cite[ASM;][]{asm96}, the Proportional Counter Array
\cite[PCA;][]{xte96} and the High-energy X-ray Timing Experiment
\cite[HEXTE;][]{hexte98} aboard \xte.  The ASM consists of three
Scanning Shadow Cameras (SSCs) sensitive to 2--10~keV photons mounted
on a rotating platform, which makes 90~s observations (``dwells'')
covering most of the sky every few hours. Data from each SSC from each
dwell are averaged to obtain the daily intensities of all known sources in
the field of view. The PCA is made up of 5 identical co-aligned
Proportional Counter Units (PCUs) sensitive to photons in the energy range
2--60~keV, and with a total area of $\approx 6000\ {\rm cm^2}$.  Data is
collected during pointed observations in standard data modes (1 and 2), as
well as a variety of user defined modes offering time resolution up to $1\
\mu s$ and 256 energy channels.  The HEXTE comprises two clusters with
$1600\ {\rm cm^2}$ total area, each of which contains 4 scintillation
detectors collimated to view a common direction and sensitive to photons
in the range 15--250~keV.

We searched for X-ray bursts in 1-s binned ``Standard-1'' mode data
from 43 public \xte/PCA observations of \src\/ between 1997 November 5
and 2000 September 27, totalling 347~ks.
We also searched in a 30~ks proprietary observation from 2002 July 27
(observation ID 70044, PI: Rothschild).
We identified 24 bursts in total.
In three cases
the PCUs were switched on during what appeared to be the decay from a
burst. We estimated the start time of these burst candidates by comparing
the decay light curve with the profile from the burst immediately before
or after. While no blackbody temperature decrease was observed in any of
these three 
cases,
no decrease was observed for the other bursts when at the same stage.
Since the observed profiles
of the other bursts are so uniform, we expect the inferred start times
to be correct to within 5--10~s. In a fourth case
the burst began while the satellite was still slewing to the source.
For that burst the measurement of the peak flux and fluence are suspect, but
the start time was precisely measured. For the analysis presented here, we
include these four bursts for calculations of recurrence times but exclude them
from any analysis involving measurements of the flux or fluence.

For each burst where we observed the entire profile we extracted PCA
spectra binned initially at 0.25~s (with progressively larger bins
towards the tail of the burst) from Event mode data (125~$\mu s$ time
resolution and 64 energy channels). We calculated a separate response
matrix for each burst using {\sc pcarsp} version 8.0 (supplied with
{\sc lheasoft} version 5.2, 2002 June 25). We fitted an absorbed
blackbody model to each spectrum, using a persistent spectrum
extracted from a 16-s interval before the burst as background.  This
approach is relatively 
\centerline{\epsfxsize=8.5cm\epsfbox{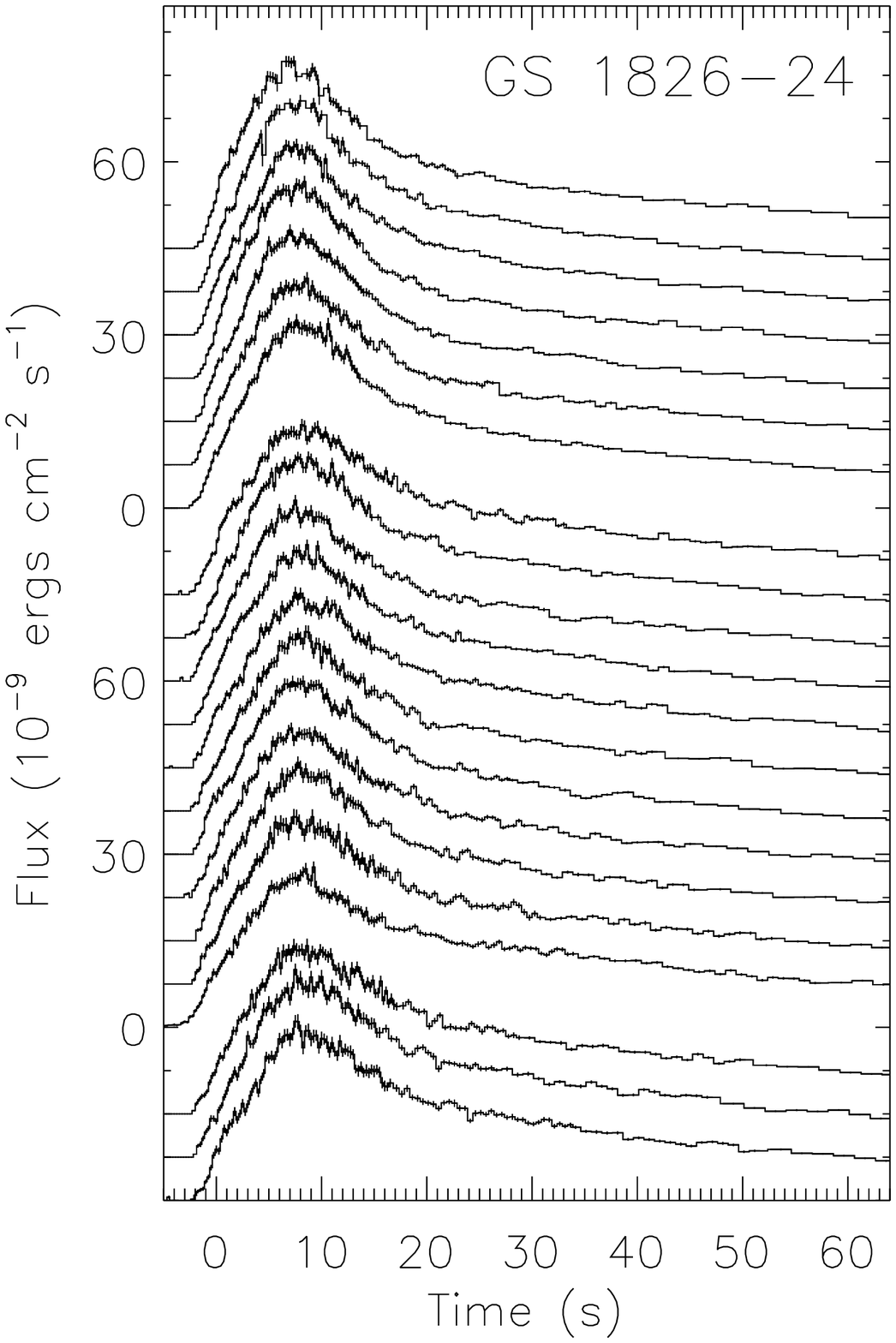}}
  \figcaption{Profiles of 20 X-ray bursts from \src\ observed by \xte\/
between 1997--2002, plotted with varying vertical offsets for clarity.  
The
upper group of 7 bursts were observed in 1997--98, the middle group
of 10 bursts in 2000, while the lower group of 3 were observed in 2002.
The bursts from each epoch have been time-aligned by
cross-correlating the first 8 seconds of the burst.
Error bars indicate the $1\sigma$ uncertainties.
  \label{profiles} }
\bigskip
\noindent
standard for X-ray burst analysis
\cite[e.g.][]{kuul01}. We estimated the
bolometric burst flux at each timestep as $F_{{\rm bol},i} =
1.0763\times10^{-11}\ T_{{\rm bb},i}^4 K_{{\rm bb},i}\ \epcs$, from
the blackbody (color) temperature $T_{{\rm bb},i}$ and normalization $
K_{{\rm bb},i}=(R_{\rm bb}/d_{\rm 10kpc})^2$, where $R_{\rm bb}$ is
the apparent photosphere radius in km, and $d_{\rm 10kpc}$ the distance to the
source in units of 10~kpc.
We define the burst start as the time at which the bolometric flux exceeds 25\%
of the peak flux, and the rise time as the interval 
for the flux to subsequently exceed 90\% of the maximum.
We fitted the flux decay to an exponential profile with
a ``break'' at which point the $e$-folding time is allowed to vary
discontinuously. We integrated the measured fluxes over $\approx150$~s
covering the burst, and extrapolated beyond this based on the exponential
fits, to derive the fluence.

We also estimated the instrumental background using {\sc pcabackest}
version 3.0 and the ``combined'' bright source models, and measured
the (absorbed) persistent 2.5--25~keV PCA flux by integrating over an
absorbed blackbody plus power law model fitted to the persistent
(pre-burst) spectra.  The mean reduced-$\chi^2$ for the persistent
spectral fits was 1.07 (56 degrees of freedom). The neutral column
density was in most cases poorly constrained and not significantly
different from zero, and in the mean was $n_{\rm
H}=(2.4\pm1.4)\times10^{22}\ {\rm cm^{-2}}$.  While this model
provided a good fit to the PCA data alone, combined fits including the
HEXTE spectrum additionally required modelling of the high-energy
spectral cutoff (see section \S\ref{secalpha}).

\section{Burst Profiles, Energetics and Recurrence Times}
\label{results}

The X-ray bursts observed by \xte\/ were remarkably similar to each
other (Fig. \ref{profiles}).
The rise times were relatively long, betseen 4.75 and 7~s 
($5.6\pm0.6$~s on average).
The first exponential decay
timescale
increased
from $14.7\pm0.7$ to $17.5\pm1.1$~s between the 1997--98 and 2000
bursts, and to $19.1\pm1.3$ for the 2002 bursts.
The variation of the burst profile with epoch is obvious in the averaged
lightcurves (Fig. \ref{profiles2}).
The second exponential timescale was, on average,
$43\pm1$~s. The peak fluxes also showed weak evidence for a decrease with
time; the mean for the 7 bursts observed in 1997--98 was
$(33.0\pm0.8)\times10^{-9}\ \epcs$, while for the bursts observed in 2000--2 it
was $(30.5\pm1.1)\times10^{-9}\ \epcs$ (note that the averages of
burst properties calculated here exclude
the bursts which we did not observe in their entirety).
This
decrease was substantially larger than the variation in the pre-burst
persistent emission (see \S\ref{inthist}, below).
Thus, it appears unlikely that the observed
variation in the peak burst flux arose as a side-effect of subtracting
the persistent emission as background.  The net effect of the
variations in peak flux and timescale was to keep the fluence
approximately constant, at $\approx1.1\times10^{-6}\ \epc$.
None of the
bursts exhibited evidence for radius expansion, so that the maximum burst
flux is a lower limit to the Eddington luminosity. 
The implied distance limit 
is consistent with that derived from previous observations.
\centerline{\epsfxsize=8.5cm\epsfbox{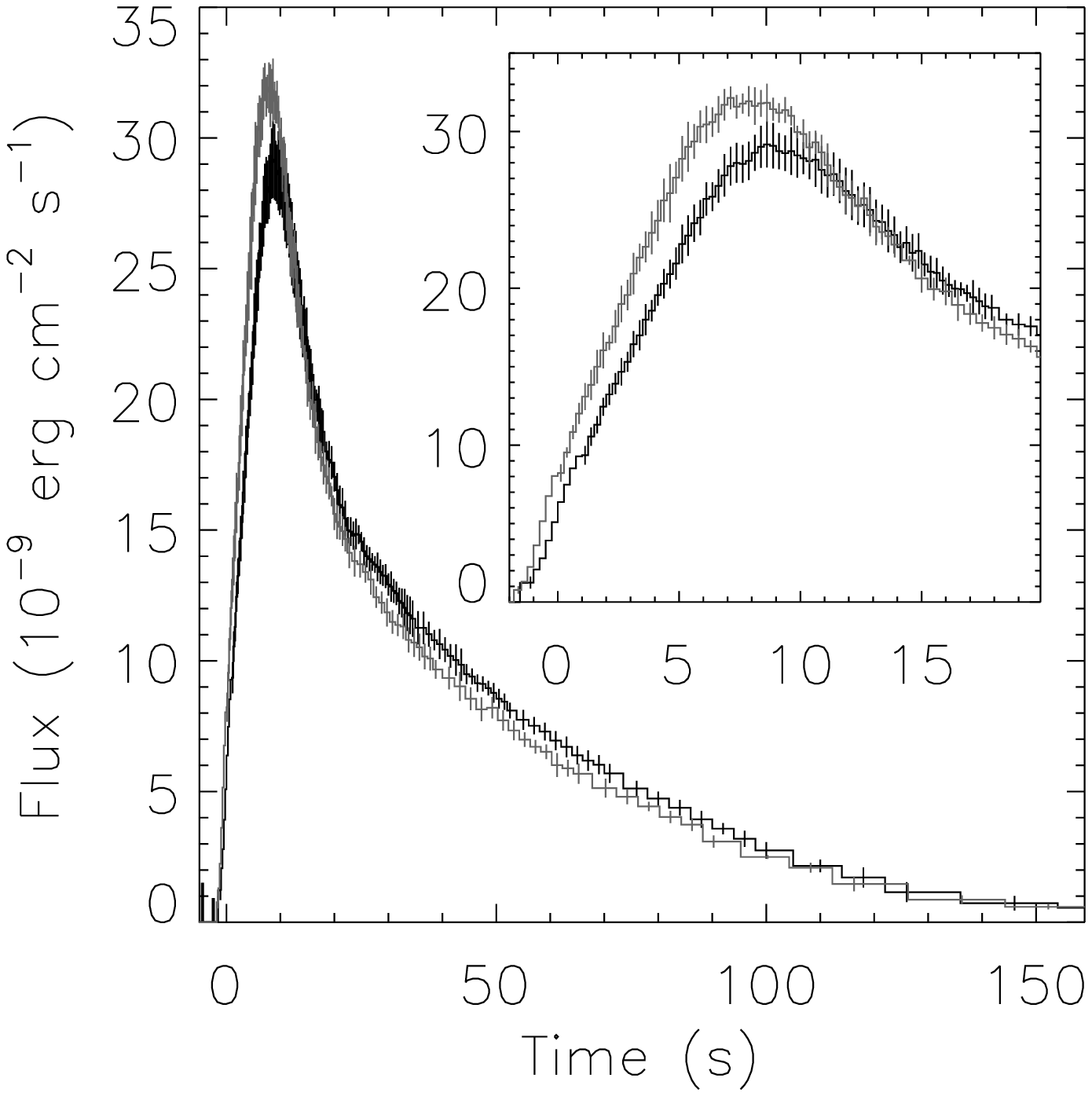}}
  \figcaption{
Mean profiles of 7 X-ray bursts from \src\ observed by \xte\/ during
1997--98 (grey histogram), and of 10 bursts observed during  2000 (black
histogram).
The bursts from 2002 have similar profiles to those from 2000.
Error bars indicate the $1\sigma$ uncertainties, derived from the scatter
of the flux within each time bin over all the bursts. The inset shows the
same profiles, expanded to show more detail around the burst rise and
peak.
  \label{profiles2} }
\bigskip
\noindent
\centerline{\epsfxsize=8.5cm\epsfbox{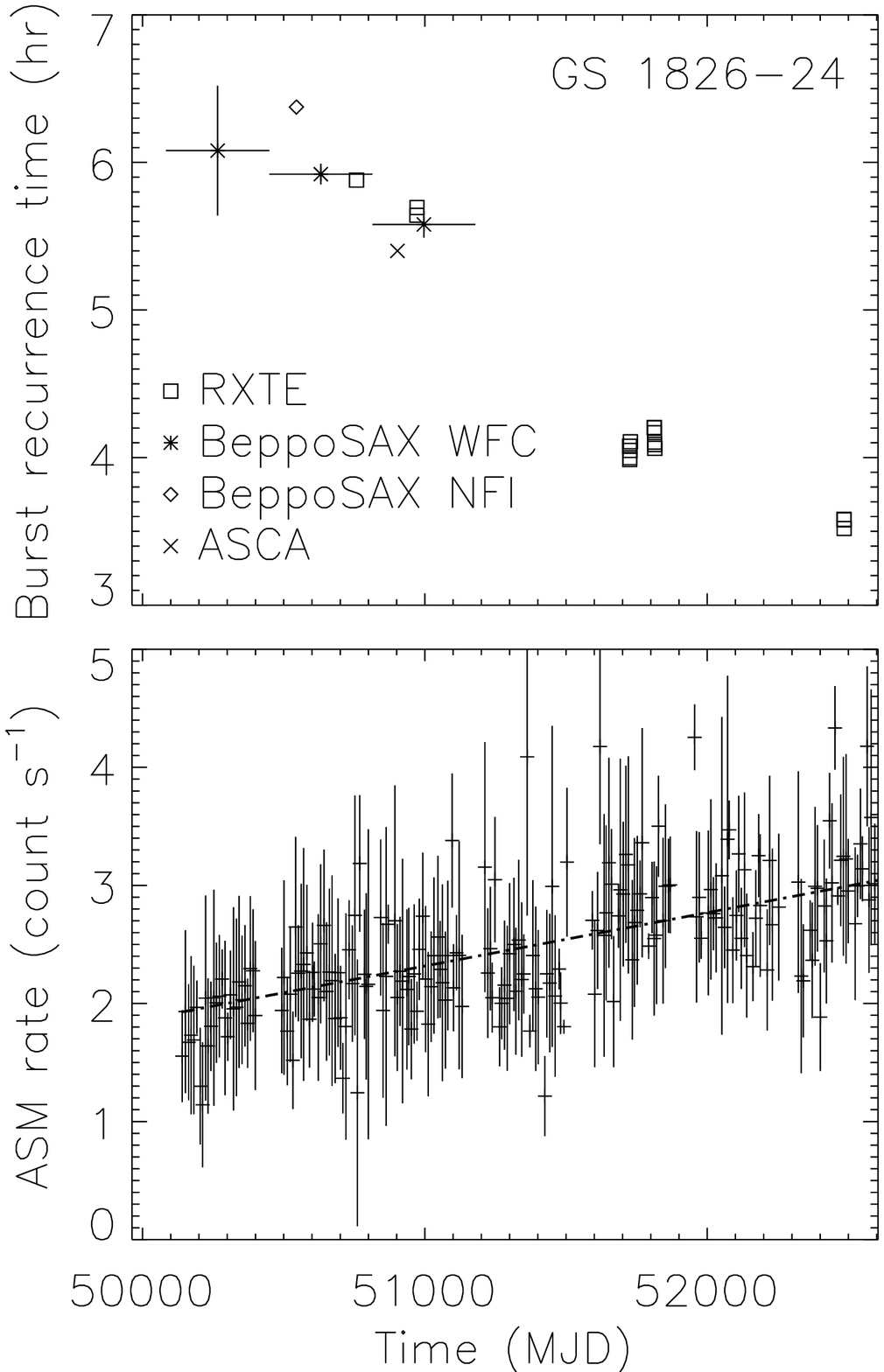}}
  \figcaption{{\it (top panel)} Recurrence time for thermonuclear
bursts from \src.  Shown are the 1997--2002 measurements by \xte\/ (open
squares), as well as earlier measurements by \sax\/
\cite[]{zand99,cocchi01c} and \asca\/ \cite[]{kong00}.
{\it (bottom panel)} Long-term ASM count rate history of \src. 10-day
averages were calculated from 1-day measurements, excluding those points
with errors $>0.7\ \cts$.  The dot-dashed line shows the linear fit curve.
The linear trend is significant at the 10$\sigma$ level.
 \label{deltat} }
\bigskip
\noindent

\subsection{Long-term burst interval history}
\label{inthist}

Due to its low Earth orbit, \xte\/ can typically only observe any given source
for 65\% of each 90~min orbit. Thus, it is likely that some
bursts were missed during the gaps between observations.  The shortest
burst intervals were found between pairs of bursts observed
on 1997 November 5--6,
2000 Jun 30, 
and 2002 July 29, at $\Delta t=5.88$, 4.00~hr, and 3.58~hr,
respectively.
While the first value is
consistent with other measurements around the same time
\cite[e.g.][]{clock99,cocchi01c}, the latter two are almost a third
shorter.
The other burst intervals measured from the \xte\/ observations are at
least a factor of 2 greater than the shortest intervals, and were
close to integer multiples of the two shortest intervals in each
epoch.  Furthermore, in each of the longer burst intervals, the
predicted intermediate burst times (assuming regular burst occurrence)
fell within data gaps. Thus, it was still possible for the bursts to
be recurring on a regular timescale.

In order to independently test for the presence of these intermediate
bursts, we examined ASM dwells during these longer intervals.  Very few
dwells occurred close to the predicted time of the bursts.
Thus, we found no evidence for bursts at the predicted intermediate times;
on the other hand, with the present data we cannot exclude the presence of
bursts within these gaps.  Taking into account these missed bursts, we
conclude that the observed burst times are
consistent with regular recurrence intervals of $5.74\pm 0.13$~hr in
1997--98,
$4.10\pm 0.08$~hr in 2000, and 
$3.56\pm0.03$~hr in 2002.

When combined with \sax\/ and \asca\/ observations between 1996--98,
as well as more recent observations with \xte,
these measurements provide evidence for a long-term, steady decrease in 
$\Delta t$ with time (Fig. \ref{deltat}).
Over the same interval, the 10-d averaged 2-10~keV ASM intensity
measurements of \src\ show evidence for a long-term trend of increasing
persistent flux.
This trend was confirmed by the PCA measurements.  The 2.5--25~keV
persistent flux measured during 1997--98 was
$(1.32\pm0.05)\times10^{-9}\ \epcs$, while in 2000 it was
$(1.80\pm0.05)\times10^{-9}\ \epcs$. This variation is significant at
the $6.4\sigma$ level.
The most recent observation, on 2002 July 29,
found the persistent 2.5--25~keV flux to be higher again, at
$(2.19\pm0.02)\times10^{-9}\ \epcs$.

\subsection{Variation of $\Delta t$ with $F_{\rm p}$}
\label{secalpha}

The burst interval measured from the \xte\/ observations decreased by
40\% between 1997--98 and 2002, while over the same period the mean
persistent flux increased by
66\%.
In order to estimate the bolometric correction to apply to the
2.5--25~keV flux, and thus estimate the ratio $\alpha$ between the integrated
persistent flux to the burst fluence
(equation \ref{ratio}), we extracted PCA and HEXTE spectra covering the
entire observation from five representative pointings
(1997 November, 1998 June, 2000 July, 2000 September and 2002 July). We fitted each
spectra to a model consisting of a Comptonisation component (``{\tt
compTT}'' in {\sc xspec}) and a Gaussian, both attenuated by neutral
absorption along the line of sight. The {\tt compTT} component
features a cutoff at around 2.8 times the temperature of the
scattering plasma, which we measured as $13.4\pm0.5$~keV in the mean.
In each case we obtained a good fit (mean $\chi^2/{\rm
dof}=1.02\pm0.13$) over the energy range 2.5--60~keV.
The mean neutral column
density was $n_{\rm H}=(1.52\pm0.25)\times10^{22}\ {\rm cm^{-2}}$,
which is consistent with the mean value derived from the PCA-only
pre-burst persistent emission fits.  We then measured the (absorbed)
flux in the 2.5--25~keV band, and compared this to the flux in the
0.1--200~keV band estimated using an idealized response
\footnote{The response used for extrapolation covered the range 0.1--200~keV
with 200 logarithmically spaced energy bins.}
in order to extrapolate outside the PCA/HEXTE bands.  We found the
bolometric correction to be $1.678\pm0.016$,
where the error represents the standard deviation from the results for
the 5 observations.

The resulting variation of the burst recurrence
interval, as a function of the bolometric persistent flux, is shown in
Fig. \ref{vstheory}; qualitatively similar results were found by
\cite{corn03a} from \sax\/ WFC observations.
For the errors on the individual $\Delta t$ measurements from \xte, we
adopted a fractional error corresponding to the standard deviation measured
by \cite{cocchi01c} of
\centerline{\epsfxsize=8.5cm\epsfbox{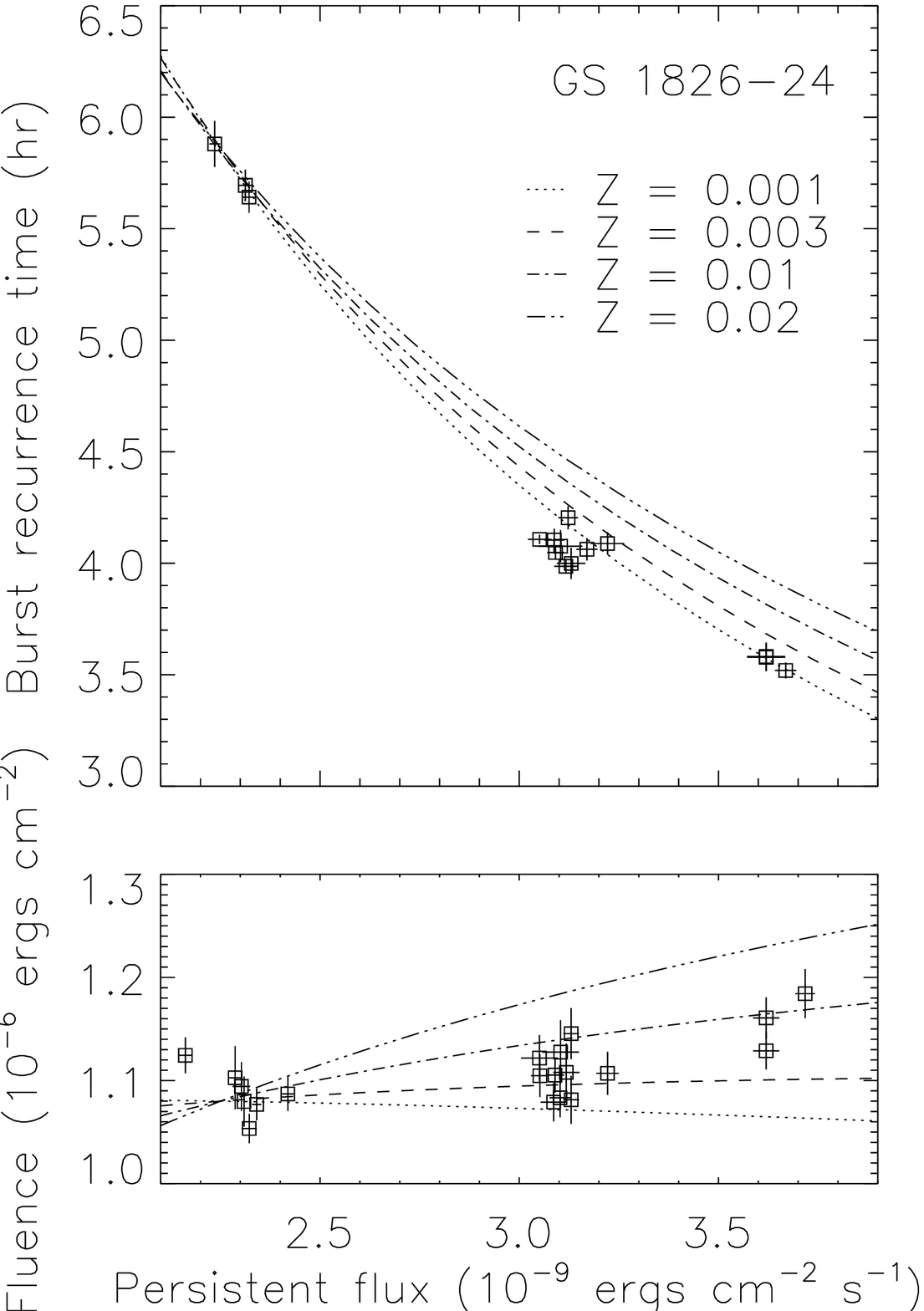}}
  \figcaption{Variation of the burst recurrence time ({\it upper
panel}) and the burst fluence ({\it lower panel}) as a function of the
estimated bolometric persistent flux in \src, from \xte\/ measurements
between 1997--2002.
Error bars indicate the $1\sigma$ errors. 
The curves show theoretical calculations for
a range of metallicities: $Z=0.02, 0.01, 0.003,$ and $0.001$. The solid
angle $(R/d)$ and gravitational energy have been chosen in each case to
match the observed fluence and recurrence time at $F_{\rm p}=2.25\times
10^{-9}\ {\rm erg\,cm^{-2}\,s^{-1}}$. For $Z=0.02, 0.01, 0.003,$ and
$0.001$, this gives $R/d=13, 10, 8, 6\ {\rm km}\ @\ 10\ {\rm kpc}$, and
$Q_{\rm grav}=175, 196, 211, 215$ MeV per nucleon.
  \label{vstheory} }
\bigskip
\noindent
0.1~hr, for a mean delay time of 5.7~hr. This
agrees well with the scatter in $\Delta t$ we observe within each epoch.
For cases where we miss $n$ or more intervening bursts, we scale 
the error
by $1/\sqrt{n+1}$, since we assume that we are observing $n+1$ burst
intervals in total. From a least-squares fit to $\Delta t$ as a function
of $F_{\rm p}$, using $\Delta t\propto F_{\rm p}^\alpha$ we found a power
law index of
$\gamma=-1.05\pm0.02$.

From the estimated bolometric fluxes we then calculated the ratio
$\alpha=E_{\rm p}/E_{\rm b}$ from each pair of bursts measured by
\xte\ (Fig. \ref{ratioplt}).
We found that $\alpha$ varied significantly with epoch, decreasing from
$\approx 44$ to $\approx40$ between 1997--98 and 2002, as the persistent
flux increased.  The weighted mean was
$41.7\pm1.6$.

From the \sax/WFC observations, \cite{clock99} derived a value of
$60\pm7$.
Roughly consistent values were determined from \sax/NFI, \xte, and
\asca\/ observations, of between 50--54 \cite[]{zand99,kong00}.
To understand the discrepancy between these measurements and the value of
$\alpha$ calculated in
the present study, we note firstly that the previous
measurements of the burst fluence from \asca\/ and \xte\/ by \cite{kong00}
agree well with our measurements.  Furthermore, all the persistent flux
measurements appear to agree in general once the difference in the energy
band is taken into account.  However, the burst fluence measured by both
the \sax\/ instruments are around 40\% lower than for \asca\ and \xte.
The \xte/PCA is known to measure fluxes that are systematically $\sim20$\%
higher than some other instruments \cite[e.g.][]{kuul02}; however, this
offset is insufficient to explain the discrepancy in the measured fluence.
Furthermore, the estimate of $\alpha$ from \xte\/ should be independent of
any systematic flux offset.
Despite the substantially higher burst fluence from the \xte\/ and \asca\/
measurements of \cite{kong00}, the calculated $\alpha$ was still close to
that measured by \sax\/ \cite[]{clock99,zand99}. This appears to result
from the bolometric correction on the 2--10~keV \xte/\asca\/ flux, which
is not quoted in the \cite{kong00} paper but we estimate at 4--6. By
comparison, the bolometric correction implied by the 2--10~keV and
0.1--200~keV \sax/NFI measurements by \cite{zand99} is 3.3.  From our
broadband spectral fits in \S\ref{secalpha}, we estimate a bolometric
correction for \xte\/ flux in the 2--10~keV band as $3.06\pm0.02$. Thus,
we attribute the higher $\alpha$ measured by \cite{kong00} to an excessive
bolometric correction factor.

\section{Comparison to Theoretical Ignition Models}
\label{comptheor}

In this section, we compare the observed burst properties with
theoretical models of type I burst ignition. We calculate ignition
conditions following \cite{cb00}, and refer the reader to that paper
for details.  Since the calculation depends only on the local vertical
structure of the layer, we give the results in terms of the local
accretion rate per unit area $\dot m$, and
the mass per unit area or column depth $y$. We assume
\centerline{\epsfxsize=8.5cm\epsfbox{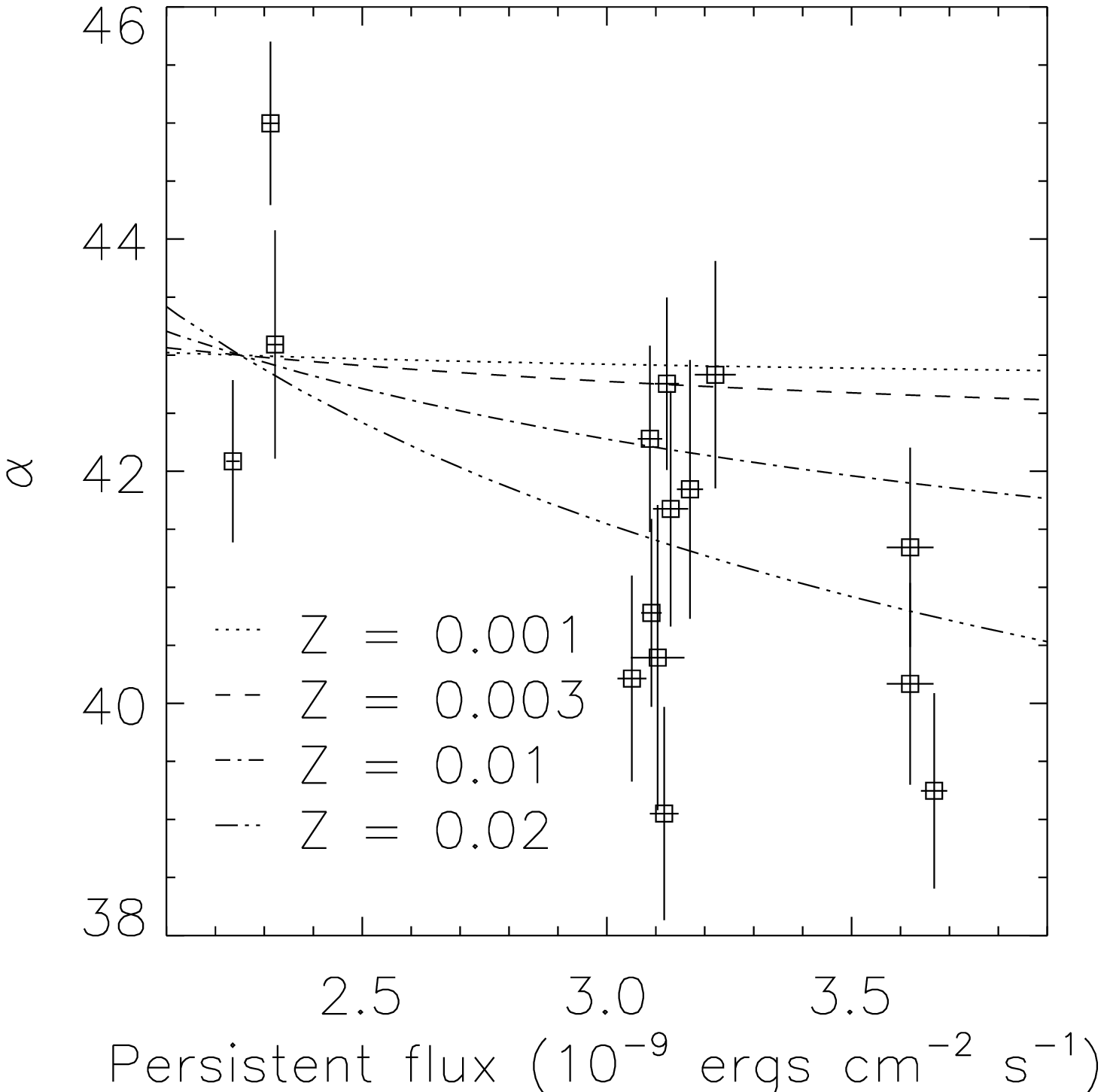}}
  \figcaption{Ratio of persistent to burst luminosity
$\alpha=L_{\rm p}/L_{\rm b}$ (equation \ref{ratio}), calculated from
\xte\/ observations between 1997 and 2002.
Error bars represent the estimated $1\sigma$ uncertainties.
The curves show theoretical calculations for the same values of
metallicity as in Fig. \ref{vstheory}.
 \label{ratioplt} }
\bigskip
\noindent
a $1.4\ M_\odot$ neutron star with radius
$R=10\ {\rm km}$, giving a surface gravity $g=(GM/R^2)(1+z)=2.45\times
10^{14}\ {\rm cm\ s^{-2}}$,
where $1+z=(1-2GM/Rc^2)^{-1/2}=1.31$ is
the gravitational redshift. This value for the redshift is close to
that recently measured for EXO 0748-676
\cite[$z\simeq0.35$;][]{cott02}.

We calculate the temperature profile of the accumulating layer of
hydrogen and helium, and adjust its thickness until a thermal runaway
occurs at the base. The temperature is mostly set by hydrogen burning
via the hot CNO cycle, and therefore the CNO mass fraction $Z$, which
we refer to as the metallicity. Our models also include compressional
heating and a flux from the crust $F_{\rm crust}$, but the results are
not sensitive to these contributions. A factor of two change in
$F_{\rm crust}$ gives a 2\% (25\%) change in ignition depth and burst
energy for $Z=0.02$ ($Z=0.001$), with a much smaller change in the
trend of these properties with $\dot m$.  We take $F_{\rm crust}$ to
be constant over the timescale of the observations, i.e. $F_{\rm
crust}=\left<\dot m\right>Q_{\rm crust}$, where the time-averaged
local accretion rate $\left<\dot m\right>$ is set equal to the value
for which the burst recurrence time is $5.7$~hr, and $Q_{\rm
crust}=0.1$ MeV per nucleon \cite[]{brown00}.

To calculate the burst energy, we assume complete burning of the H/He
fuel layer, and that the accreted material covers the whole surface of
the star. The total energy is then $4\pi R^2 y Q_{\rm nuc}\,\xi_{\rm
b}^{-1}/(1+z)$, where $y$ is the ignition column depth, $Q_{\rm nuc}$ is
the energy per gram from nuclear burning, and the factor $\xi_{\rm b}$
accounts for anisotropic burst emission. We set the latter parameter
equal to unity initially, but use it to track where burst anisotropy
might play a role. We write $Q_{\rm nuc}=1.6+4\left<X\right>$ MeV per
nucleon, where $\left<X\right>$ is the mass-weighted mean hydrogen
fraction at ignition. This expression for $Q_{\rm nuc}$ assumes
$\approx 35$\% energy loss due to neutrinos during the rp process
\cite[e.g.][]{fuji87}, and gives $4.4$ MeV per nucleon for
approximately solar hydrogen abundance ($X=0.7$).

Table \ref{theorytbl} gives ignition conditions for four different
metallicities. In each case, we adjust $\dot m$ until
$\Delta t=(y/\dot m)(1+z)=5.7\ {\rm hr}$,
as observed in 1997--98. Lower metallicity models have
less hot CNO heating during accumulation, giving a larger ignition
column and therefore larger $\dot m$ to match $\Delta t=5.7\ {\rm
hr}$. We are free to vary $\dot m$ because of the uncertainty in
the $F_p$--$\dot m$ relation. The ratio of observed flux and inferred
$\dot m$ constrains a combination of the gravitational energy release
per nucleon, and the degree of anisotropy in the persistent or burst
emission. The persistent flux is
\begin{equation}\label{eq:flux}
F_{\rm p}={L_{\rm X}\over 4\pi d^2}={\dot m Q_{\rm grav}\over 1+z}
\left({R\over d}\right)^2 \xi_{\rm p}^{-1},
\end{equation}
where $\xi_{\rm p}$ accounts for anisotropy in the persistent
emission, and $Q_{\rm grav}=c^2z/(1+z)\approx GM/R$ is the
gravitational energy release per gram. The burst fluence is
\begin{equation}\label{eq:fluence}
E_b={Q_{\rm nuc}\,y\over 1+z}\left({R\over d}\right)^2\xi_{\rm
b}^{-1}.
\end{equation}
Inserting equations (\ref{eq:flux}) and (\ref{eq:fluence}) into
$\alpha=F_{\rm p}\Delta t/E_b$ gives
\begin{equation}\label{eq:alpha}
\alpha\,Q_{\rm nuc}=\left({\xi_{\rm b}\over\xi_{\rm
p}}\right)c^2z=290\ {\rm MeV}\ \left({z\over
0.31}\right)\left({\xi_{\rm b}\over\xi_{\rm p}}\right).
\end{equation}
The last column in Table \ref{theorytbl} gives $\alpha Q_{\rm nuc}$
for the models. Reconciling these values with equation
(\ref{eq:alpha}) requires $\xi_{\rm b}/\xi_{\rm p}=0.5$--$0.6$, within
the range of expected values (e.g.~\citealt{lap85,fuji88}), especially
since the ``anisotropy'' factor for the persistent emission includes
other effects such as energy radiated outside the X-ray band.

The predicted variations in fluence, recurrence time, and $\alpha$
with $\dot M$ are compared to the observations in Figures
\ref{vstheory} and \ref{ratioplt}. We adjust each curve to match the
observed fluence, recurrence time, and $\alpha$ at $F_{\rm
p}=2.25\times 10^{-9}\ {\rm erg\ cm^{-2}\ s^{-1}}$.

The model with $Z=0.02$ (approximately equivalent to solar metallicity)
agrees very well with the
observed burst energy and variation in $\alpha$ with $\dot M$. The
predicted burst energy (Table \ref{theorytbl}) is
$5.3\times 10^{39}\ {\rm ergs}$,
corresponding almost exactly to the observed fluence for a
distance of 6 kpc. The $\approx 10$\% variation in $\alpha$ arises
because hydrogen burning occurs between bursts as the layer
accumulates. In a given fluid element, the time to burn all the
hydrogen via the beta-limited hot CNO cycle is $11\ {\rm hr}\
(Z/0.02)^{-1}(X_0/0.7)$ (e.g.~\citealt{bil98a}), where $Z$ is the CNO
mass fraction, and $X_0$ is the accreted hydrogen fraction. This means
that as the recurrence time decreases from $5.7$ to $4$~hr,
significantly more hydrogen per unit mass is present when the burst
ignites, leading to an increase in $Q_{\rm nuc}$, and corresponding
decrease in $\alpha$. Although our models do not address the time
evolution of the burst, the variation in mean burst lightcurves shown
in Fig.~\ref{profiles2} is consistent with a change in the fuel
composition.

However, the solar metallicity model is less successful at explaining
the observed variations in $\Delta t$ and $E_{\rm b}$
(Fig.~\ref{vstheory}). Whereas the observations show a decreasing
ignition mass, $\Delta t\propto\dot M^{-1.05}$, and a $\approx 3$\%
rms variation in $E_{\rm b}$, the models have $\Delta t$ decreasing
less steeply than $1/\dot M$, and a $\approx 10$\% increase in $E_{\rm
b}$ with $\dot M$. These variations arise because as recurrence time
drops, less hydrogen burning occurs, less helium is present at
ignition, and the ignition mass increases.

Extra heating of the accumulating layer would give a reduced ignition
mass and better agreement with the observed trends of $\Delta t$ and
$E_{\rm b}$. For the $Z=0.02$ model in Table \ref{theorytbl}, the
extra flux required to match the observed change in $\Delta t$ is
$\approx 0.5$ MeV per nucleon for an increase of 50\% in $\dot
M$. This could perhaps be provided by residual heat from the ashes of
previous bursts \cite[]{taam93,woos03}, although time-dependent
simulations are required to test this.

A second possibility is that the fraction of the neutron star surface
covered by accreted fuel changes with $\dot M$, as suggested by
Bildsten (2000). We can infer the covering fraction from measurements
of the blackbody normalization in the tail of the bursts.
We show in Fig.\ \ref{radius} the mean value of
$R_{\rm bb}/d_{\rm 10kpc}$ between $\approx20$--50~s after the start
of each burst. We staggered the time window over which each average
was calculated, to ensure that the mean blackbody temperature over
each time interval was approximately constant.  We found a significant
decrease of $\approx20$\% in the blackbody normalization. A variation
of the covering fraction by this amount, and the resulting increase in
local accretion rate $\dot m$, would go some way to bringing the
$\Delta t$ and $E_{\rm b}$ variations into better agreement. However,
the detailed relationship between the measured $R_{\rm bb}$ and the
covering fraction is uncertain.

\centerline{\epsfxsize=8.5cm\epsfbox{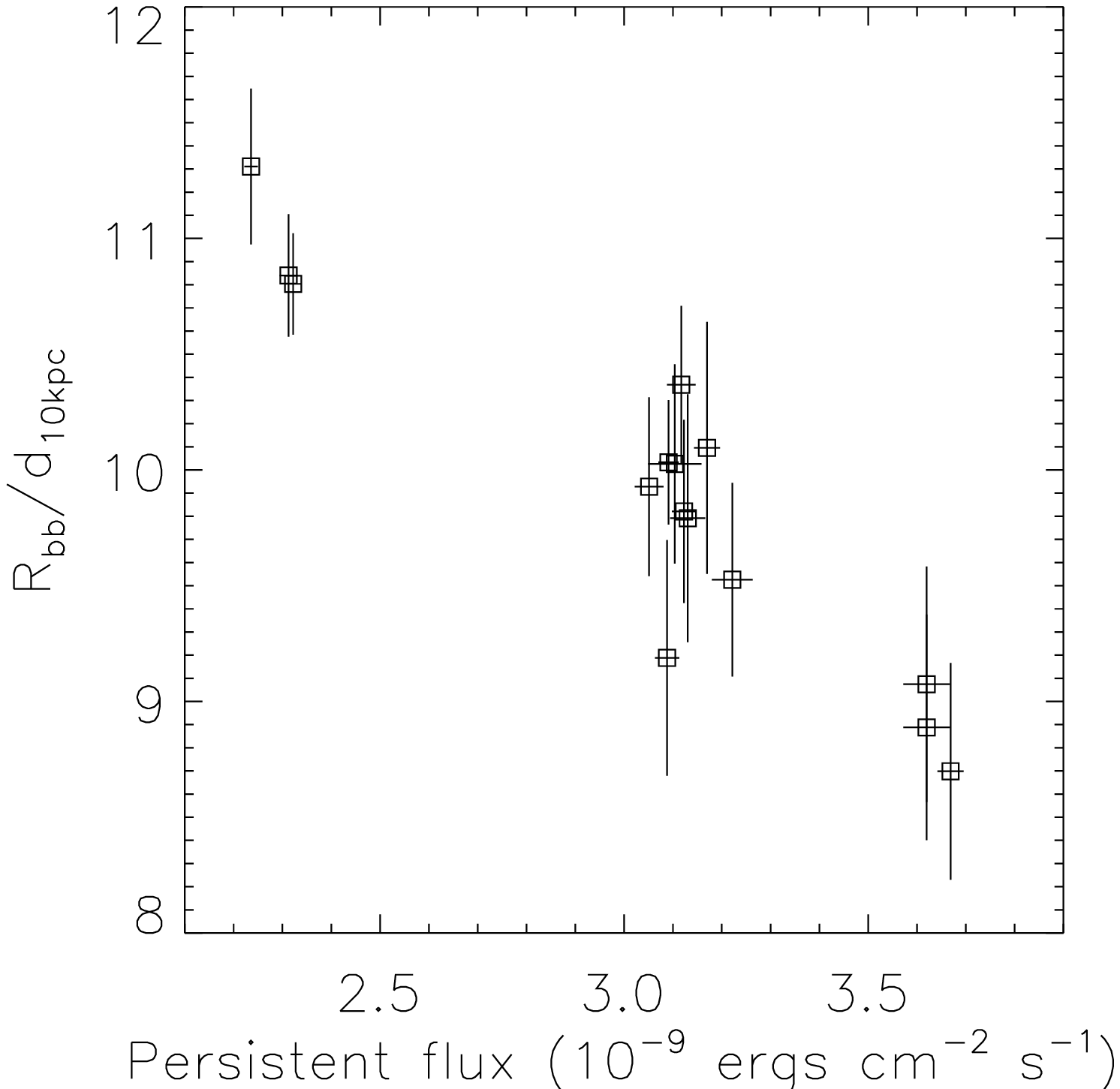}}
 \figcaption{Blackbody normalization $R_{\rm bb}/d_{\rm 10kpc}$ as
a function of bolometric persistent flux, for bursts from \src. The
normalization was averaged over a time window $\approx20$--50~s following
the start of the burst. Error bars indicate the $1\sigma$ uncertainties.
\label{radius} }
\bigskip
\noindent

While the observed variation in $\alpha$ argues strongly against low
metallicity models (which have a constant composition at ignition
since little hot CNO burning occurs during accumulation),
such models give much better
agreement with the observed trends in $\Delta t$ and $E_{\rm b}$. The
constant composition leads to an ignition mass almost 
independent of
$\dot M$, giving a constant burst fluence and $\Delta t$ variations
close to $1/\dot M$. Low metallicity models overpredict the burst
energy, since the reduced CNO heating leads to a larger ignition
mass. Accounting for this discrepancy by increasing the distance to
the source requires $d\approx 10$--$12\ {\rm kpc}$, above the current
upper limit. However, there are other possible ways to bring the burst
energy into agreement. Recent time-dependent simulations of bursts by
\cite{woos03} show that leftover hydrogen and CNO nuclei in the ashes
of the previous burst lead to extra heating, reducing the critical
mass. Also, the fuel may cover only part of the neutron star
surface. Table \ref{theorytbl} lists the value of $\xi_{\rm
b}^{-1/2}(R/d)$ that gives agreement with the observed fluence
(eq.~[\ref{eq:fluence}]), where we now interpret $R$ as a measure of
the area covered by accreted material rather than the neutron star
radius. For the low metallicity models, we find $\xi_{\rm
b}^{-1/2}R\approx 5$--$6$~km at 6 kpc, or covering fractions (assuming
a neutron star radius of 10~km) of $30$--$40$\%.

Finally, we address the sensitivity of our results to the fraction of
accreted hydrogen. Recent evolutionary models involving intermediate
mass binaries \cite[]{prp02} propose that the secondary has undergone
some main sequence hydrogen burning, leading to a relatively
hydrogen-poor donor ($X\approx 0.35$--$0.7$). As an example, we show
results for an accreted hydrogen mass fraction $X_0=0.5$ in the lower
half of Table \ref{theorytbl}. The ignition conditions are not very
sensitive to the total amount of hydrogen, since only a small amount
of hydrogen burns during accumulation. However, the burst energy is
significantly reduced, since $Q_{\rm nuc}$ depends strongly on the
amount of hydrogen versus helium. The variation in burst properties
with flux is not a useful way to distinguish the amount of hydrogen,
and is almost identical with Figures \ref{vstheory} and \ref{ratioplt}
for the $X=0.5$ models.

\section{Discussion}

The regularity of the bursting from \src\ makes it a unique source to
compare burst theory and observations. In this paper, we have
presented {\it RXTE}/PCA observations of 24 type I X-ray bursts from
GS~1826-238. For the first time for this source, we have been able to
study the properties of bursts in detail at different mass accretion
rates. The bursts from a given epoch have identical lightcurves
(Fig. 1 and 2), and show long tails likely powered by rp-process
hydrogen burning, as suggested by Bildsten (2000). The bursts from a
given epoch are consistent with a single recurrence time, and this
varies with accretion rate as $\Delta t\propto\dot{M}^{-1.05}$. The
burst fluence increased slightly with the accretion rate, and $\alpha$
decreased by $\approx 10$\%, with mean value
$41.7\pm1.6$.

The decrease in $\alpha$ with $\dot M$ implies that stable burning of
hydrogen occurs between bursts. This requires solar metallicity
($Z\approx 0.02$) in the accreted layer (Fig.~\ref{ratioplt}). Solar
metallicity models also give good agreement with the observed burst
energies, as pointed out by Bildsten (2000). However, the small
variation in fluence, and $\Delta t$ variation with $\dot M$ is
difficult to explain with solar metallicity models, which predict
$10$\% variations in fluence, and $\Delta t$ decreasing less steeply
than $1/\dot M$. Low metallicity models ($Z\sim 0.001$) naturally
explain the small variation in burst fluence, and $\approx 1/\dot M$
scaling of $\Delta t$, since little CNO burning occurs between bursts
giving an ignition mass almost independent of $\dot M$. However, the
observed $\alpha$ variations with $\dot M$ rule out low metallicity
models.

There are several possible ways to reconcile the solar metallicity
models with the observations of $\Delta t$ and $E_{\rm b}$. Studies of
timing and spectra of LMXBs indicate that $L_X$ is not always a good
indicator of $\dot M$ (e.g.~\citealt{vdk90}), we have assumed
$L_X\propto\dot M$ here. Extra heating of the accumulating layer would
act to reduce the critical mass and bring the observations and theory
into agreement. One possibility is that residual heat from the ashes
of previous bursts heats the layer (Taam et al.~1993; Woosley et
al.~2003), time-dependent simulations are needed to test this. If the
fraction of the neutron star surface covered by fuel changes with
$\dot M$, the changing local accretion rate per unit area could also
reconcile the models and observations. We find that the blackbody
radius $R_{\rm bb}$ in the tail of the bursts decreased by $\approx $20\%
between the observed epochs. If this indicates a change in covering
fraction, it would almost be enough to explain the discrepancy. However,
the covering fraction decrease with $\dot M$ is opposite to the increase
suggested by Bildsten (2000) to explain trends in burst properties.
Furthermore, $R_{\rm bb}$ is at best an uncertain measure of the
emitting area. Due to variations in the atmospheric opacity with photon
energy, the measured blackbody temperature is generally higher than the
effective temperature \cite[e.g.][]{lth86,sth03}. Consequently, $R_{\rm
bb}$ is expected to give an underestimate of the emitting area, by a
factor which depends on the detailed structure and composition of the NS
atmosphere.
Better $R/d$ measurements could be obtained with high resolution spectra
and appropriate model atmospheres.

\cite{woos03} have recently calculated time-dependent multi-zone burst
models including a large nuclear reaction network. They show that both
the rise and decay times of the burst depend sensitively on nuclear
properties in the rp-process path, and in addition find burst rise times
of several seconds, in good agreement with the observations. There is
much to learn from a detailed comparison of time-dependent simulations
with the burst lightcurves. In particular, the variation in decay time
with accretion rate may directly reflect the changing abundance of
hydrogen at ignition. 

Many bursters show a transition from regular, frequent bursts with
long durations at low $\dot m$, to irregular, infrequent and short
bursts at high $\dot m$, a behavior first discovered by {\it EXOSAT}
\cite[]{vppl88}. By comparing {\it BeppoSAX}/WFC observations of nine
regular type I bursters, \cite{corn03a} propose that this behavior is
common to all bursters, and that the transition occurs at a universal
luminosity $L_X\approx 2\times 10^{37}\ {\rm erg\ s^{-1}}$. While the
physics of the transition to irregular bursting is not understood
(e.g.~see discussion by Bildsten 2000), we have shown in this paper
that the regular bursting seen in \src\ is well-understood as being
due to helium ignition in a hydrogen-rich environment. This argues
that other bursters such as KS~1731-260 are in this burning regime
when regular bursting is seen (and not bursting via unstable hydrogen
ignition, as suggested by van Paradijs et al.~1988 and Cornelisse et
al.~2003). {\it BeppoSAX} found a peak burst rate for KS~1731-260 (the
burster with the largest range in observed $\dot m$'s) of $\approx 9$
bursts per day. Extrapolating the observed $1/\dot M$ dependence of
$\Delta t$, we expect \src\ to reach this burst rate, and perhaps
transition to irregular bursting at $F_p\approx 4.6\times 10^{-9}\
{\rm erg\ cm^{-2}\ s^{-1}}$.  Further monitoring of the burst behavior
of \src\ will hopefully probe the physics of the transition to
irregular bursting.

\acknowledgments

This research has made use of data obtained through the High Energy
Astrophysics Science Archive Research Center Online Service, provided
by the NASA/Goddard Space Flight Center.
This work was supported in part by NASA under grants NAG 5-9184 and NAG
5-8658, contract NAS5-30720 and by the National Science Foundation under
grant PHY99-07949.  L.  B. is a Cottrell Scholar of the Research
Corporation. A.C. is supported by NASA Hubble Fellowship grant HF-01138
awarded by the Space Telescope Science Institute, which is operated for
NASA by the Association of Universities for Research in Astronomy, Inc.
under contract NAS 5-26555.



\clearpage

\begin{deluxetable}{lcccllcccc}
\tabletypesize{\scriptsize}
\tablecaption{Ignition conditions for bursts with  $\Delta t=5.7\ {\rm hours}$\tablenotemark{a}\label{theorytbl}}
\tablewidth{0pt}
\tablehead{
\colhead{$Z$} & \colhead{$\dot m_4$} & \colhead{$y_8$} & \colhead{$T_8$} &
\colhead{$X$} & \colhead{$\left<X\right>$} & \colhead{$Q_{\rm nuc}$\tablenotemark{b}} 
& \colhead{$E_{\rm burst}$\tablenotemark{c}} &
\colhead{$R/d$\tablenotemark{d}} & \colhead{$\alpha Q_{\rm nuc}$\tablenotemark{e}}\\
\colhead{} & \colhead{} & \colhead{} & \colhead{} &
\colhead{} & \colhead{} & \colhead{} &\colhead{$(10^{39}\ {\rm erg})$} & \colhead{$($km @ $6$ kpc$)$} & \colhead{
}
}
\startdata
\multicolumn{10}{c}{Accreted H fraction $X_0=0.7$}\\
\hline
0.02 & 0.97 & 1.5 & 2.2 & 0.40 & 0.55 & 3.8 & 5.3 & 9.6 & 152\\
0.01 & 1.4 & 2.1 & 2.1 & 0.55 & 0.62 & 4.1 & 8.0 & 7.8 & 164\\
0.003 & 2.1 & 3.3 & 2.0 & 0.65 & 0.68 & 4.3 & 13 & 6.1 & 172\\
0.001 & 2.8 & 4.3 & 1.9 & 0.68 & 0.69 & 4.4 & 17 & 5.3 & 175\\
\hline
\multicolumn{10}{c}{Accreted H fraction $X_0=0.5$}\\
\hline
0.02 & 0.92 & 1.4 & 2.1 & 0.20 & 0.35 & 3.0 & 4.0 & 11 & 120\\
0.01 & 1.2 & 1.9 & 2.0 & 0.35 & 0.42 & 3.3 & 5.8 & 9.2 & 132\\
0.003 & 1.8 & 2.9 & 1.9 & 0.45 & 0.48 & 3.5 & 9.3 & 7.3 & 140\\
0.001 & 2.4 & 3.8 & 1.8 & 0.48 & 0.49 & 3.6 & 12 & 6.3 & 143\\
\enddata
\tablecomments{The first six columns are: CNO mass fraction $Z$; local
rest mass accretion rate at the surface of the star $\dot m_4=\dot m/10^4\ {\rm g\ cm^{-2}\ s^{-1}}$;
ignition column depth $y_8=y/10^8\ {\rm g\ cm^{-2}}$; ignition
temperature $T_8=T/10^8\ {\rm K}$; hydrogen mass fraction at the base
$X$; mean hydrogen mass fraction $\left<X\right>=\int dy X(y)/y$.}
\tablenotetext{a}{$\Delta t=(y/\dot m)(1+z)$ is the recurrence time measured by an observer at infinity.}
\tablenotetext{b}{Nuclear energy release, $Q_{\rm nuc}=1.6+4.0\left<X\right>$ MeV per nucleon.}
\tablenotetext{c}{Predicted burst energy assuming the fuel covers the surface of the star, $E_{\rm burst}=4\pi R^
2yQ_{\rm nuc}\xi_{\rm b}^{-1}/(1+z)$ (we take $R=10\ {\rm km}$,
$z=0.31$, $\xi_{\rm b}=1$). The observed burst energy is $4.9\times
10^{39}\ {\rm erg}\ (d/6\ {\rm kpc})^2$.}
\tablenotetext{d}{The $R/d$ that gives agreement with the observed burst fluence. This quantity is $\xi_{\rm b}^{
-1/2}(R/d)$ if the emission is anisotropic.}
\tablenotetext{e}{We adopt the observed value $\alpha=40$.}
\end{deluxetable}

\end{document}